\documentclass[sigconf]{acmart}
\usepackage{caption}
\usepackage{subcaption}
\usepackage{color,soul}
\AtBeginDocument{%
  }

\setcopyright{iw3c2w3}

\acmConference[HAIPS @ CCS 2025]{Human-centered Evaluation and Auditing of Language Models Workshop at ACM CCS}{October 17, 2025}{Taipei, Taiwan}

\acmDOI{}

\acmISBN{}

\begin{document}

\title{Beyond Permissions: Investigating Mobile Personalization with Simulated Personas}


\author{Ibrahim Khalilov}
\email{ikhalil1@jh.edu}
\affiliation{
  \institution{Johns Hopkins University}
  \city{Baltimore}
  \state{Maryland}
  \country{USA}
}

\author{Chaoran Chen}
\email{cchen25@nd.edu}
\affiliation{%
  \institution{University of Notre Dame}
  \city{Notre Dame}
  \state{Indiana}
  \country{USA}
}

\author{Ziang Xiao}
\email{ziang.xiao@jhu.edu}
\affiliation{
  \institution{Johns Hopkins University}
  \city{Baltimore}
  \state{Maryland}
  \country{USA}
}

\author{Tianshi Li}
\email{tia.li@northeastern.edu}
\affiliation{%
  \institution{Northeastern University}
  \city{Boston}
  \state{Massachusetts}
  \country{USA}
}

\author{Toby Jia-Jun Li}
\email{toby.j.li@nd.edu}
\affiliation{%
  \institution{University of Notre Dame}
  \city{Notre Dame}
  \state{Indiana}
  \country{USA}
}

\author{Yaxing Yao}
\email{yaxing@jhu.edu}
\affiliation{
  \institution{Johns Hopkins University}
  \city{Baltimore}
  \state{Maryland}
  \country{USA}
}


\renewcommand{\shortauthors}{Khalilov, et al.}

\begin{abstract}
Mobile applications increasingly rely on sensor data to infer user context and deliver personalized experiences. Yet, the mechanisms behind this personalization remain opaque to users and researchers alike. This paper presents a sandbox system that uses sensor spoofing and persona simulation to audit and visualize how mobile apps respond to inferred behaviors. Rather than treating spoofing as adversarial, we demonstrate its use as a tool for behavioral transparency and user empowerment. Our system injects multi-sensor profiles—generated from structured, lifestyle-based personas—into Android devices in real time, enabling users to observe app responses to contexts such as high activity, location shifts, or time-of-day changes. With automated screenshot capture and GPT-4 Vision-based UI summarization, our pipeline helps document subtle personalization cues. Preliminary findings show measurable app adaptations across fitness, e-commerce, and everyday service apps such as weather and navigation. We offer this toolkit as a foundation for privacy-enhancing technologies and user-facing transparency interventions.
\end{abstract}





\maketitle

\section{Introduction}

Mobile applications are deeply embedded in daily life, enabling navigation, social networking, and personalized services. These conveniences, however, come at the cost of continuous and often opaque data collection. Apps routinely access GPS location, sensor readings, microphone inputs, browsing activity, and other system data, creating complex data flows that users rarely comprehend. For example, a weather app might log location data every few minutes, even when not in use, and a sports app may collect users' movement patterns or Bluetooth signals to infer nearby devices \cite{lu2023detecting, AlmuhimediHazim2015YLhb}. These practices have raised increasing concerns in recent media and research~\cite{NYT2018}. However, despite users' privacy concerns, their behaviors often contradict these concerns, which is generally known as the \textit{privacy paradox}~\cite{baruh2017big, kokolakis2017privacy}.

Compared to desktop web tracking, mobile applications have deeper, real-time access to sensitive data, often in ways that appear harmless but create unexpected risks when combined with third-party services. For instance, granting GPS access to a navigation app may seem reasonable, yet embedded advertising networks can repurpose this access to continuously track users’ movements~\cite{embedded_GPS}. Likewise, microphone permissions intended for voice commands can enable unintended background audio collection, raising surveillance concerns \cite{Gao2019}. A \textit{New York Times} investigation found that at least 75 companies collected and monetized precise location data from millions of mobile devices in the U.S., often without users’ clear understanding or consent \cite{NYT2018}. These examples illustrate how seemingly innocent permissions can lead to unforeseen privacy risks through multi-modal data collection in mobile apps.

Yet, general users often have very limited means to understand the complexity of data collection and the associated privacy risks in mobile systems \cite{church2015understanding, capponi2019survey, beierle2020data}. While platforms offer several mechanisms (e.g.,privacy policies, Android’s Data Safety section, Apple’s App Privacy labels, etc.), these tools are often unusable in practice: they tend to be overly long, vague, hard to interpret, or disconnected from meaningful context \cite{ObarandOeldorf-Hirsch2020, privacy_label, privacy_nutrition_labels, LinYanzi2023DSvA}. Recent research has shown that even newer interventions like iOS privacy labels fail to significantly improve users’ understanding or ability to make informed choices \cite{yao2022usable, LinYanzi2023DSvA}. As a result of these shortcomings, users are often left with little real choice: they frequently consent to data collection not because they accept the terms, but because denying permissions means sacrificing access or functionality. In other words, the limitations of current mechanisms not only undermine awareness, but also create environments where even privacy-conscious users struggle to anticipate the downstream effects of their decisions, resulting in consent that is more procedural than genuinely informed \cite{capponi2019survey, beierle2020data}.

These limitations in user understanding have deeper implications when real-time data is used not just for collection, but for shaping user experiences directly. Modern mobile personalization systems compound these issues by operating behind the scenes, relying on continuous real-time sensor data, ranging from GPS location to accelerometer motion and ambient light, to infer users’ behavioral contexts and dynamically adjust app content \cite{delgado2022survey, harari2016using}. While these features enable adaptive experiences, they raise further concerns about transparency and user agency, particularly as users are seldom aware of when sensors are active or how behavioral data is interpreted by apps \cite{kroger2022privacy}. These mechanisms often function invisibly, making it difficult to trace how environmental signals or personal routines shape the digital experiences users encounter \cite{acquisti}. As such, the shift toward sensor-driven personalization adds another layer of opacity that users are not well-equipped to manage.

Conventional Privacy-Enhancing Technologies (PETs) such as location blurring, sensor noise injection, or access restriction frameworks (e.g., PrivaSense~\cite{privaSense}) mainly aim to limit the data available to apps. Even though, these tools reduce exposure, they rarely help users understand how personalization decisions are actually made~\cite{cha2018privacy}. Prior work shows that even small or seemingly harmless sensor readings can be combined to reveal sensitive behaviors~\cite{malekzadeh2018protecting}, a process often described as the behavioral inference pipeline~\cite{ataiefard2021deep}. In mobile environments, where sensing is continuous and signals from multiple sensors are fused, this pipeline creates privacy risks that remain largely invisible to users~\cite{delgado2022survey}.

Taken together, the evolving privacy risks in mobile systems and users' limited understanding of how their data contribute to these risks constitute a critical gap in the literature. Prior work has shown that part of this gap stems from the opacity of data flows and users’ hesitation to experiment with privacy settings due to fear of breaking functionality or exposing sensitive information~\cite{chen2024empathy}. To address this, recent research has proposed privacy sandbox environments, a safe, controlled space where users can explore how data inputs influence system behaviors without risking real-world consequences~\cite{chen2024empathy}. These sandboxes create opportunities for experimentation, education, and greater transparency by allowing users to test privacy-relevant scenarios without exposing personal data or altering persistent system settings.

Building on this idea, we introduce a novel LLM-based mobile sandbox that enables users to interactively explore the connection between mobile sensor data and associated privacy implications in a risk-free environment. Our system allows users to simulate mobile app interactions using synthetic, persona-driven sensor data and observe how apps adapt to different behavioral contexts, which helps users build more concrete mental models of how personalization and profiling mechanisms work.

To evaluate the feasibility of this sandbox approach, we developed a real-time sensor spoofing toolkit that replaces live sensor data with structured, simulated user profiles that reflect different lifestyle patterns (e.g., an active commuter, a sedentary worker, or a frequent traveler). Using this prototype, we conducted an initial experiment to test whether and how mobile apps respond to synthetic data. Our results suggested that the synthetic data successfully tricked various mobile apps, allowing the apps to respond to the fed data (e.g., fitness apps award activity badges without physical movement, shopping apps localize content based on spoofed GPS, and weather apps dynamically adjust UI and forecast based on time-of-day spoofing). 


These results inform the design of an interactive sandbox that allows users to explore how varying types of sensor data influence app behaviors. Our investigation is guided by the following research questions:

\begin{itemize}
\item \textbf{RQ1:} \label{rq1} What visible changes do users experience in mobile app behavior when synthetic, context-specific sensor data is introduced?
\item \textbf{RQ2:} \label{rq2} How can relatable personas help users make sense of these changes?
\end{itemize}

\begin{itemize}
\item \textbf{RQ3:} \label{rq3} How might a sandbox-based toolkit help users understand mobile personalization and support transparency-focused privacy practices?
\end{itemize}

In this project, we also aim to examine whether such a system can enhance user understanding of how contextual sensor data shape app behavior, and whether that understanding could translate into greater trust, agency, or willingness to engage with behavioral transparency tools. In addition, we seek to identify which features of the system are most effective in supporting comprehension and enabling meaningful exploration of mobile personalization.

At a high level, our approach seeks a new paradigm for users to actively explore and reflect on the transparency of mobile ecosystem. This ongoing work aims to define a novel PET design space which encourages users' experimentation and interpretation of how behavioral data influences app logic. This work makes the following contributions: 

\begin{itemize}
    \item A working prototype for mobile persona simulation that leverages Frida \cite{frida}, LSposed \cite{lsposed}, and the Motion Emulator app \cite{reed2024zhufucdev} on rooted Android devices to automate multi-sensor spoofing in real-time.
    \item Empirical findings demonstrating that a variety of apps, including shopping platforms, fitness trackers, and utility apps, respond dynamically and measurably to contextual changes introduced by sensor spoofing.
    \item A conceptual design for a user-facing sandbox that allows individuals to select a persona, activate spoofing conditions, and observe how apps respond to behavioral inputs from alternative user profiles.
    \item A roadmap for system extensions, including GPT-based persona generation, GPT-4 Vision-based UI summarization, synthetic Google Calendar input, and instrumentation of network/API behavior to support behavioral transparency and auditability.
\end{itemize}

While still in development, our system demonstrates potential to support future tools for user education, transparency audits, and mobile behavior research. Ultimately, we propose that giving users the means to simulate and observe their digital self shaped by sensor data can open new avenues for privacy awareness and accountability in data-driven personalization systems.

\section{Related Work}

Privacy risks in mobile ecosystems have grown substantially with the rise of sensor-rich devices and ubiquitous background data collection. Numerous studies emphasize that users often lack visibility into how their behavioral data is collected, inferred, and used by applications \cite{masood2022tracking, weiss2012impact}. For instance, mobile apps regularly access accelerometer, GPS, and light sensors to build granular user profiles and personalize recommendations or ads \cite{masood2022tracking}. Research into ``digital phenotyping'' has shown how sensor data is repurposed to monitor user mood and mental health \cite{onnela2016harnessing}, raising serious ethical and privacy concerns. Yet, tools that allow users to meaningfully observe or intervene in this personalization pipeline remain limited \cite{weiss2012impact, ebrahimi2021mobile}.

To address this, privacy-enhancing technologies (PETs) have traditionally focused on static protections like anonymization or minimization. However, recent efforts emphasize transparency and user-side experimentation. \citet{xian2025user, aaraj2024vbit} explore visual analytics and sandboxing for user experimentation, while \citet{ayalon2017developers} examine how developers balance privacy design with user experience. Still, many of these tools require technical expertise or are limited to controlled use cases. On-device protection mechanisms such as those proposed by \citet{malekzadeh2018protecting} mitigate sensitive inferences by transforming raw sensor data before sharing. Similarly, \citet{narain2018mitigating} introduced PrivoScope, which provides synthetic GPS trajectories to help users track and manage location-based app behaviors. These efforts improve observability but generally treat individual sensors in isolation, without integrating broader behavioral context or multiple modalities.

Recent research has also turned toward frameworks that help users make sense of their data through interactive feedback. \citet{chen2024ccs} introduced a system that allows users to audit personalized web recommendations and understand algorithmic logic, aligning with broader transparency goals. In the mobile space, however, reverse engineering tools like Frida \cite{frida} and LSposed \cite{lsposed} remain mostly targeted at technical users, and require scripting knowledge for app testing or spoofing scenarios. As such, the opportunity to democratize these tools for broader privacy exploration is still largely untapped.

\citet{chen2024empathy} propose an empathy-based sandbox to help users understand how data influences web experiences, showing how interactive exploration can bridge the gap between privacy attitudes and behaviors. Our work builds on this foundation and applies it to the mobile domain, where the stakes are often higher due to sensor-rich tracking. By allowing users to simulate realistic behavioral personas using synthetic, sensor-driven data, we offer a low-risk method for studying mobile personalization dynamics without exposing users' own data. This complements existing work on privacy behaviors while introducing a novel integration of persona simulation, multi-sensor spoofing, and visual UI inspection to surface hidden personalization mechanisms.

To operationalize these ideas, we present a two-part contribution: (1) a system that enables users to simulate behavioral patterns through real-time sensor spoofing using structured personas, and (2) an experiment that allows users to observe how mobile apps visibly adapt to these simulated contexts.

\section{Methodology}

We built a prototype system that simulates user behaviors through a combination of persona generation, sensor spoofing, and visual analysis. This section outlines our current implementation and technical components.

\subsection{Personas in Design}

Personas play a critical role in privacy research by making abstract risks more tangible \cite{chen2024empathy, chen2024ccs}. Rather than focusing solely on abstract sensor values, our system grounds those values in recognizable human behaviors and demographics. By tying spoofed sensor data to narrative user profiles,
we give structure to what would otherwise be invisible personalization mechanisms. This framing helps both researchers and users reason about when app behavior aligns or misaligns with their expectations.

\begin{figure}[ht]
\centering
\begin{minipage}[c]{0.35\linewidth}
\centering
\includegraphics[width=\linewidth]{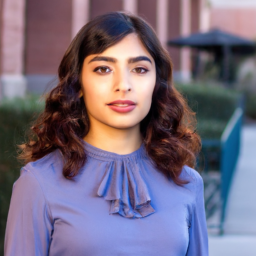}
\end{minipage}
\hfill
\begin{minipage}[c]{0.6\linewidth}
\centering
\includegraphics[width=\linewidth]{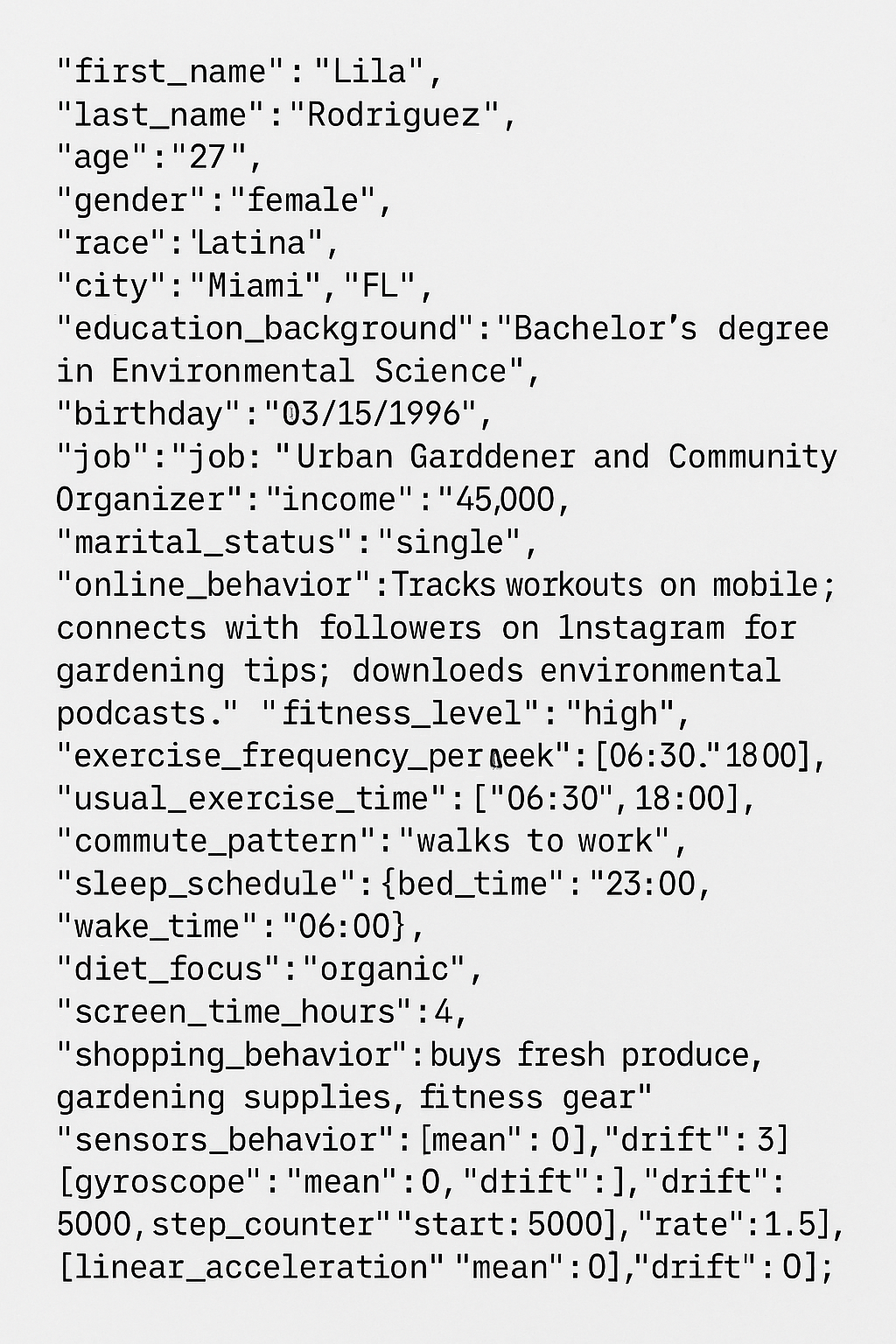}
\Description{Profile image and JSON-style demographic and behavioral traits for Lila Rodriguez.}
\end{minipage}
\caption{Simulated persona of Lila Rodriguez. Left: profile image. Right: JSON-style demographic and behavioral traits used to generate sensor spoofing patterns for mobile personalization evaluation.}
\label{fig:lila_profile_side}
\end{figure}

We have already implemented a robust persona generation pipeline that draws on language models to create diverse, context-rich profiles. One such persona, Lila Rodriguez, is a 27-year-old Latina who works as a community organizer and urban gardener. She frequently uses her mobile phone to track runs, browse sustainable living content, and discover local farmers' markets. Her behavioral profile reflects a moderate-to-high fitness level, plant-based diet, and daily outdoor routines, all of which are mapped to spoofed sensor traits like early-morning light exposure, frequent step activity, and elevated motion during commute hours. Figure~\ref{fig:lila_profile_side} presents Lila's demographic attributes and synthesized portrait, illustrating how grounded, human-readable profiles guide both sensor mapping and interface evaluation.

These profiles are designed not only for realism but for internal consistency, which capture how someone like Lila, who walks or bikes to work and practices yoga in the evening, might appear through motion and system data. Her profile, generated through GPT-4 via OpenAI's API, includes demographic context, lifestyle traits, and sensor mappings such as increased drift in accelerometer and step counter values during active hours.

Moving forward, we continue refining these personas to more closely match real-world variation in behavior and environment. We aim to expand into areas like browser profile spoofing (e.g., simulating Chrome history and Google ad IDs) to reflect more of a user's digital footprint. Our approach draws inspiration from prior research showing that rich, data-driven personas can foster user empathy and improve engagement with privacy decisions \cite{hrynenko2024identifying}, but shifts the focus toward active, sensor-based experimentation rather than static educational tools.

By anchoring our spoofing approach in personas like Lila's, we create a more interpretable layer for detecting behavioral inference and identifying mismatches between spoofed behavior and app response. This approach offers a practical bridge between low-level data spoofing techniques and the broader, user-facing implications of app personalization.

\subsection{Persona Generation and Sensor Mapping}

Our pipeline begins with generating diverse user personas using GPT-4. Each persona is designed to reflect a realistic lifestyle by combining demographic attributes (e.g., age, job, location), behavioral routines (e.g., exercise frequency, screen time, commuting patterns), and structured sensor traits. These include distributions for physical activity (e.g., accelerometer and step count), environmental context (e.g., light, magnetic field), and temporal characteristics (e.g., typical wake/sleep times, exercise hours).

While we use the term ``persona'' for clarity and alignment with prior work in privacy and HCI \cite{chen2024empathy}, our implementation extends beyond narrative user profiles. Each persona functions as a parameterized behavioral agent, which encodes structured sensor-level patterns that directly drive spoofing inputs. This dual nature means they act both as interpretable lifestyle narratives for human reasoning and as executable behavioral models for the system.

We ground our persona design in established persona methodologies from HCI and privacy research, which use rich demographic and behavioral narratives to support usability evaluation, and model realistic user contexts~\cite{privaSense, delgado2022survey, harari2016using, kroger2022privacy, acquisti, chen-etal-2025-towards-design}. Prior work by Chen et al.~\cite{chen2024empathy} applied persona-based approaches to study privacy reasoning in browser interactions; however, their scope was limited to web browsing behaviors and did not integrate sensor-level simulation. Our work extends this approach to the mobile domain, where multi-sensor inputs (e.g., accelerometer drift, ambient light variation, GPS mobility patterns, etc.) can be systematically parameterized and injected into real devices to drive app behavior.

\textbf{Persona design process.} Each persona is generated by sending a structured prompt-based request to GPT-4 using a standardized template that specifies:
\begin{enumerate}
    \item \textbf{Demographics:} age, gender, location, occupation, and income brackets.
    \item \textbf{Lifestyle patterns:} commuting habits, daily mobility range, exercise frequency, and typical app use times.
    \item \textbf{Sensor behavior parameters:} statistical ranges for motion, light, magnetic field, and temporal activity patterns, mapped from the lifestyle attributes.
    \item \textbf{Environmental context:} urban vs. rural lighting patterns, weather influence on motion, and indoor/outdoor time distribution.
\end{enumerate}

The model returns a detailed profile with these attributes plus a corresponding JSON mapping that defines the persona's ``digital footprint'' (e.g., expected accelerometer variance, daily step rate, light exposure curves). We implement validation constraints in the generation script to ensure plausibility, such as preventing night-shift workers from having high morning activity, or avoiding unrealistic GPS movement speeds.

For example, Carlos Ramirez, a 25-year-old software developer in Austin, exhibits high screen time, low physical activity, and late evening mobile usage, mapped to low-movement sensors but elevated light readings. Another persona, Linda Johnson, a 45-year-old nurse with moderate fitness routines and daytime mobile use, maps to elevated motion and light values in morning hours.

This approach ensures consistency while allowing for diverse, context-rich personas that can simulate realistic usage patterns. The persona output also includes a synthetic profile image and short lifestyle summary to support interpretability. Similar persona-based methodologies have been shown in prior work to increase empathy and improve users' ability to reason about mobile data privacy from another person's perspective \cite{chen2024empathy}. In our case, the structured personas help guide both the sensor spoofing inputs and the interpretation of app responses.

\subsection{Sensor Spoofing Infrastructure}

To simulate persona-driven sensor environments, we leverage the Motion Emulator app \cite{reed2024zhufucdev}, an LSposed-based module designed for rooted Android devices. Our experimental setup consists of a Magisk-rooted \cite{magisk} Android phone running LSposed, integrated with a locally hosted Frida server for real-time instrumentation. We execute Frida commands directly through the Termux \cite{termux} terminal on the device to launch, hook, and manipulate the Motion Emulator application during runtime. A custom-built interface feeds structured sensor data into the emulator, allowing us to inject temporally synchronized spoofed values while the emulator records input across different sensors.

\begin{figure*}
\centering
\includegraphics[width=0.90\linewidth]{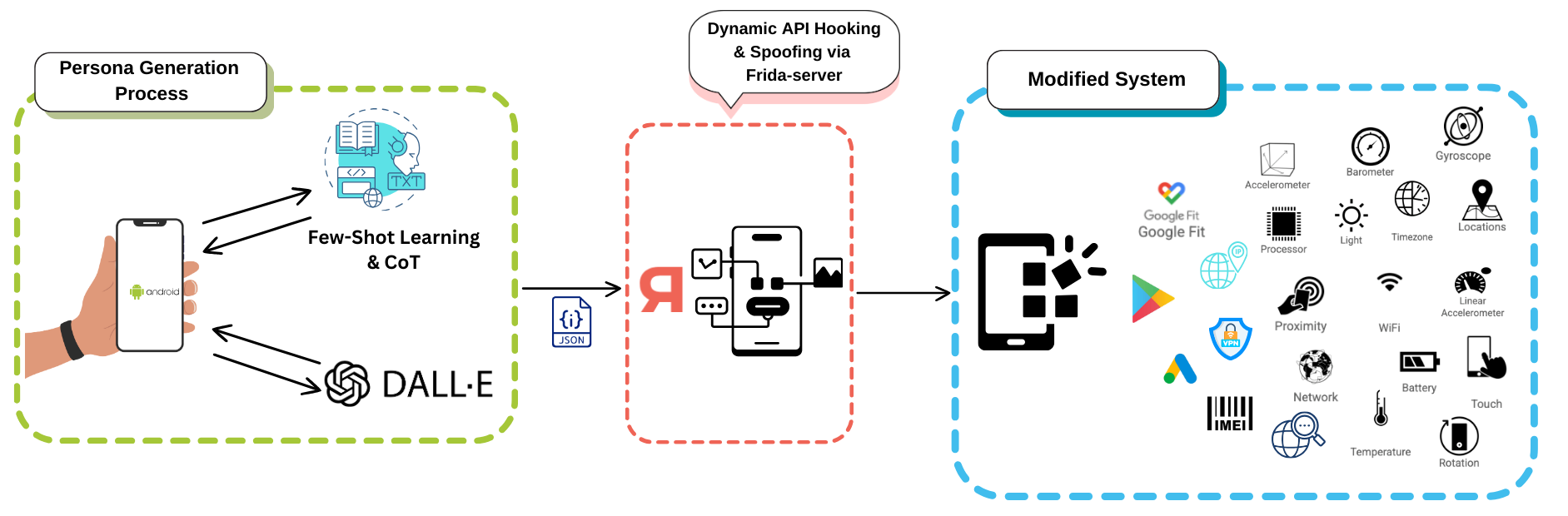}
\captionsetup{font=footnotesize}
\caption{Current pipeline for Persona Generation and Real-Time Data Spoofing.}
\label{fig:pipeline}
\vspace{-0.5em}
\end{figure*}

The system currently supports spoofing a wide range of behavioral and environmental signals, including accelerometer, gyroscope, linear acceleration, ambient light, step counter, step detector, rotation vector, gravity, magnetic field, orientation, GPS location, cell tower station, system time, and time zone. Once injected, these values are relayed through the Android sensor subsystem, allowing mobile applications to process them as though they originated from genuine user behavior. As a result, apps that rely on these contextual signals dynamically respond to the simulated conditions, e.g., changing their interface layouts, triggering different content modules, or adjusting interaction flows in ways consistent with the persona being emulated  (see Figure~\ref{fig:pipeline}).

\subsection{Automation and Visual Monitoring}

To observe app responses in realistic usage scenarios, we developed a lightweight automation layer that simulates typical user behavior. Each session begins with launching a suite of commonly used mobile apps, including Facebook, Spotify, Uber, and a weather app, alongside one or two target applications selected for observation. These apps were chosen to reflect common mobile routines, spanning social interaction, media consumption, navigation, and environmental updates, as supported by empirical studies showing these categories dominate daily smartphone use patterns across time and context \cite{ferreira2011understanding}. The automation scripts replicate familiar usage sequences such as browsing social media, listening to music, navigating, and lightly switching between apps. During the entire session, persona-driven sensor values are injected continuously in the background, shaping the behavioral context.

Rather than following a rigid procedure, our system emulates a fluid and realistic usage environment where multiple apps are active under spoofed conditions. The Motion Emulator is activated to inject persona-specific sensor values across the entire device, affecting not only the target app (e.g., Etsy or Weather) but also background apps such as Spotify, Facebook, or Uber. These background apps are not merely for ambiance; they are part of the simulated behavior ecosystem and also respond to the spoofed sensor data. Throughout each session, timed screenshots are captured at predefined intervals to record how interfaces across various apps evolve under the influence of the simulated behavioral context. We chose to trigger screenshots shortly after each app launch, using slightly randomized delays, to better reflect how users typically experience app content without creating a rigid or artificial usage pattern. This allows us to collect interface snapshots that feel natural and varied, rather than narrowly scripted.

To analyze how user interface elements change under these simulated behavioral conditions, we employ GPT-4 Vision \cite{gpt4} to generate natural language summaries of each screenshot. These summaries extract visible content such as banners, product cards, notifications, and time-sensitive elements. A follow-up GPT-4 prompt compares pairs of screenshots to detect changes in layout, recommendations, or presented content, highlighting any influence introduced by spoofed sensor data.

By combining persona-grounded app activity with structured visual summarization, our system approximates how a real user's behavior may shape mobile app experiences. This enables systematic observation of personalization mechanisms that are typically opaque, offering a clearer window into how behavioral contexts influence application behavior.


Together, these components form the basis of a sandbox environment for user-driven experimentation with mobile personalization dynamics.
The system enables users to actively probe hidden personalization mechanisms by controlling how their device ``appears'' through spoofed sensor inputs. By shifting spoofing from a circumvention tactic to a transparency tool, the sandbox allows users to explore questions such as: \textit{How does this app respond if I appear highly active in the morning?} or \textit{What changes occur when I mimic someone who browses primarily at night or commutes frequently during the day?} These scenarios are grounded in structured lifestyle simulations, such as a bakery owner managing early deliveries or a sedentary tech worker active late at night, with each mapped to temporally coordinated sensor patterns. This hands-on approach helps users observe how behavioral cues shape their mobile experience, fostering greater awareness of underlying personalization systems.

\subsection{Threat Model and Scope}
Our current sandbox focuses on personalization mechanisms that manifest as client-side, UI-visible changes triggered by real-time sensor data. These include adaptations such as location-based product recommendations, time-of-day–dependent interface changes, and motion-triggered fitness badges. By limiting our scope to effects that appear directly in the user interface during an active session, we can systematically observe and document changes without requiring backend access or invasive instrumentation.

We do not, at present, address personalization processes that occur entirely in server-side systems or that require long-term behavioral profiling, such as latent user inference, targeted ads pipelines, or cross-platform tracking. In future iterations, we plan to expand the system to capture and analyze network-level signals, including DNS activity, to better understand how apps communicate in response to different behavioral contexts. This will be complemented by the integration of an agentic system capable of autonomously probing apps, logging responses, and linking observed UI changes with underlying communication patterns.

\subsection{Ethical Considerations and Reproducibility}
While our system is intended for research and educational purposes, its underlying techniques, particularly sensor spoofing and persona-based simulation, could be misused for activities such as evading fraud detection, manipulating location-based services, or fabricating behavioral patterns (e.g., falsifying driving data to influence insurance premiums). Although many commercial platforms employ safeguards to detect such manipulation, the risk of enabling harmful uses remains.

To mitigate this risk, we will not publicly release the complete, production-ready codebase. Instead, we plan to share a non-exploitable subset of resources, including persona generation templates, anonymized example personas, and partial automation scripts, only with verified researchers who have a documented, legitimate research purpose.

Because GPT-generated personas may also introduce hallucination risks, such as implausible life histories or unrealistic behavioral patterns, we incorporate multiple safeguards before deployment. These include prompt constraints that enforce internal consistency, plausibility checks to verify that demographic and behavioral attributes align with the intended persona scenario, and manual review by the research team to screen for harmful or nonsensical profiles. This vetting process reduces the likelihood of introducing unrealistic personas that could distort experimental results or simulate unsafe behaviors.

This approach balances responsible disclosure with reproducibility. While the full system requires a rooted Android device, Frida, and LSposed, the released resources will allow others to reproduce key aspects of our process, adapt the methodology to their own contexts, and validate findings without enabling malicious or unsafe deployments.

\section{Preliminary Results}

Our early experiments demonstrate that injecting different persona-driven sensor contexts leads to measurable app adaptations: fitness apps award activity badges without physical movement, shopping apps localize content based on spoofed GPS, and weather apps dynamically adjust UI and forecast based on time-of-day spoofing. These consistent behavioral shifts indicate that sensor-based profiling is not only active but observable and reproducible. While our system is still in development, these findings suggest its potential to support future tools for user education, transparency audits, and mobile behavior research. We contribute a technical prototype, a repeatable testing pipeline, and a design rationale for repurposing sensor spoofing as a foundation for behavioral transparency and user-centered analysis in mobile ecosystems.

We tested our system on over ten Android applications across categories such as e-commerce, fitness, navigation, and utilities, and observed clear signs that sensor-driven personalization is both active and detectable. Our evaluations simulated diverse behavioral contexts by spoofing GPS location, system time, motion sensors, and light exposure, providing a window into how different apps respond to manipulated user environments.

\begin{figure*}
    \centering

    \begin{minipage}[t]{0.30\linewidth}
        \centering
        \includegraphics[height=0.40\textheight]{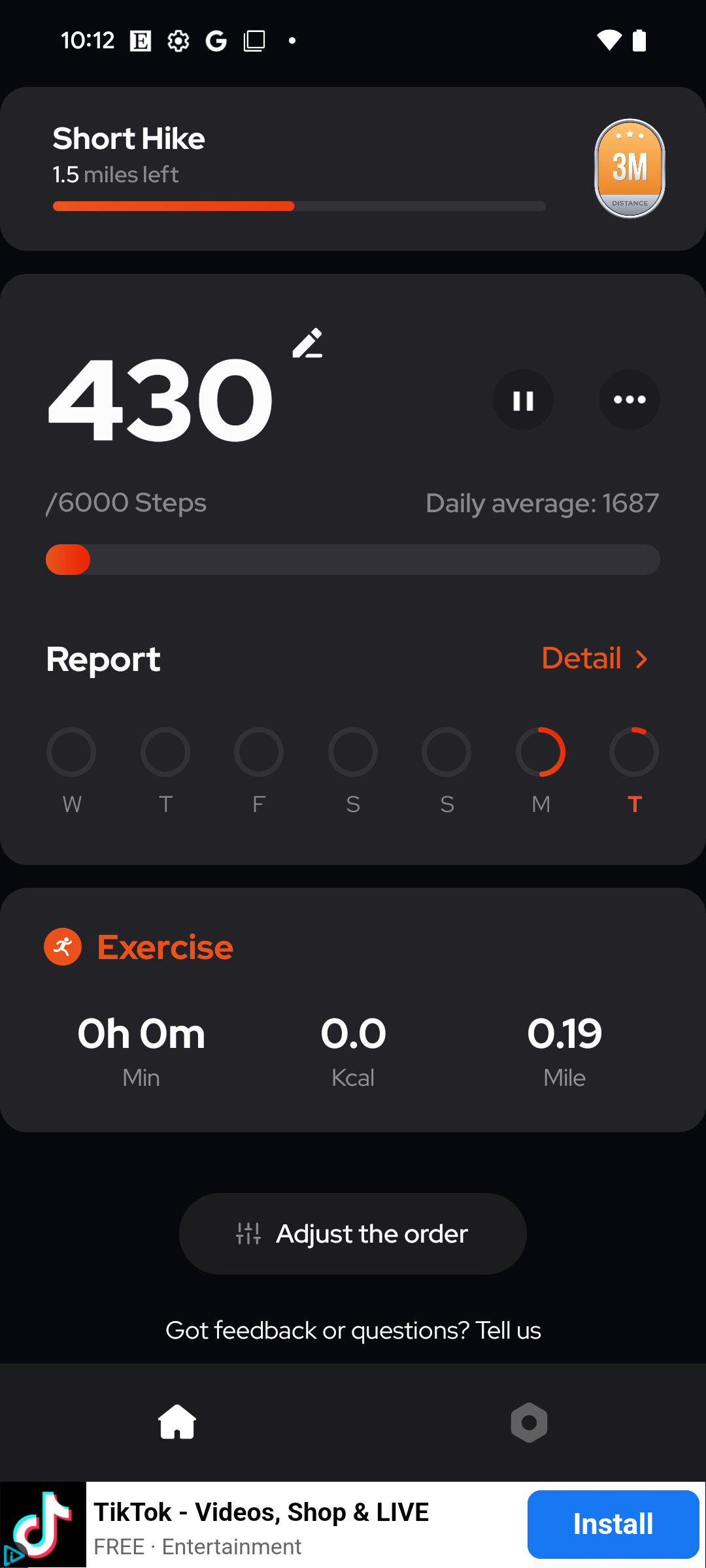}
        \captionsetup{font=footnotesize}
        \subcaption{Simulated high step activity for the "Fitness Enthusiast" persona. Step counter spoofed to emulate frequent walking and trigger fitness tracking behavior.}
        \label{fig:3a}
    \end{minipage}
    \hfill
    \begin{minipage}[t]{0.30\linewidth}
        \centering
        \includegraphics[height=0.40\textheight]{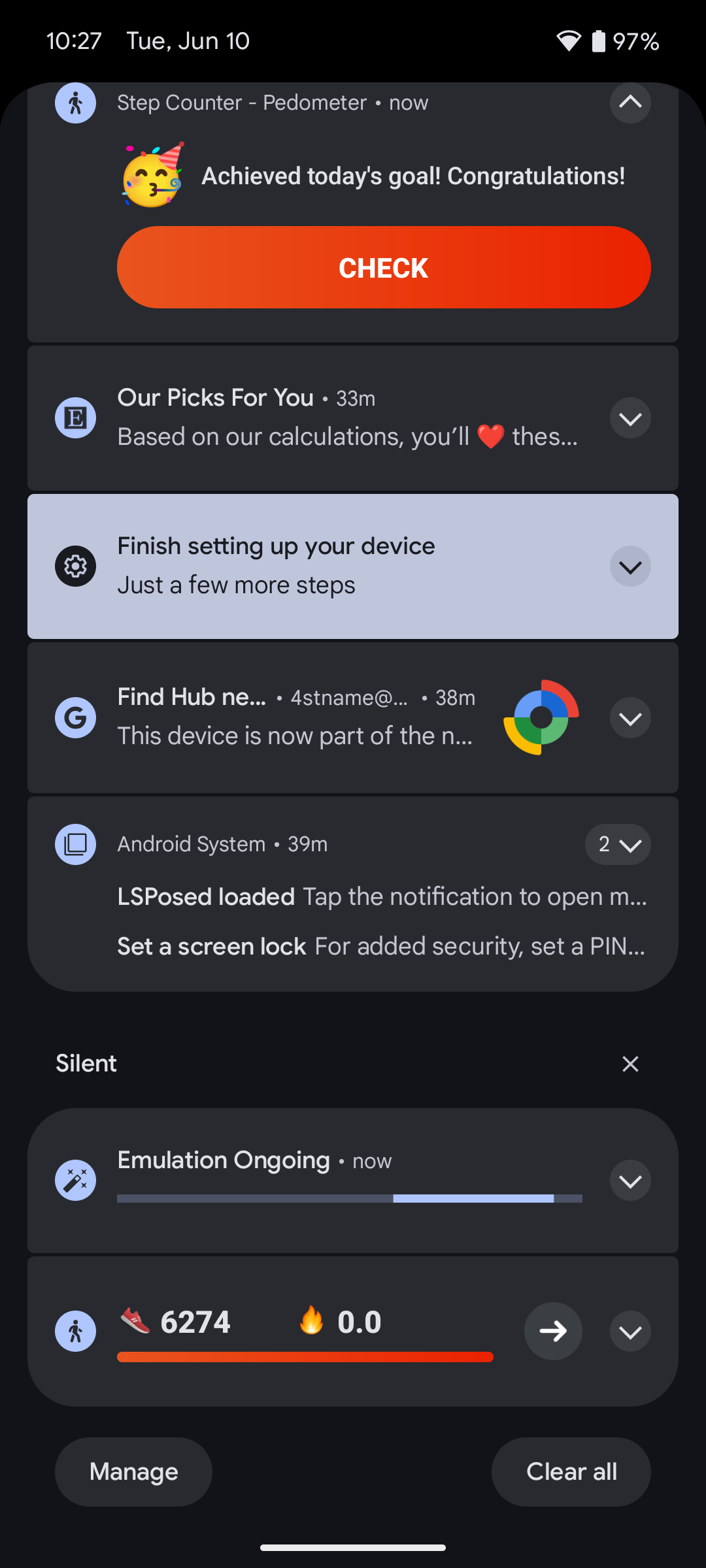}
        \captionsetup{font=footnotesize}
        \subcaption{Fitness app reacts to spoofed step data by issuing a reward badge, demonstrating behavior change in response to spoofed physical activity.}
        \label{fig:3b}
    \end{minipage}
    \hfill
    \begin{minipage}[t]{0.30\linewidth}
        \centering
        \includegraphics[height=0.40\textheight]{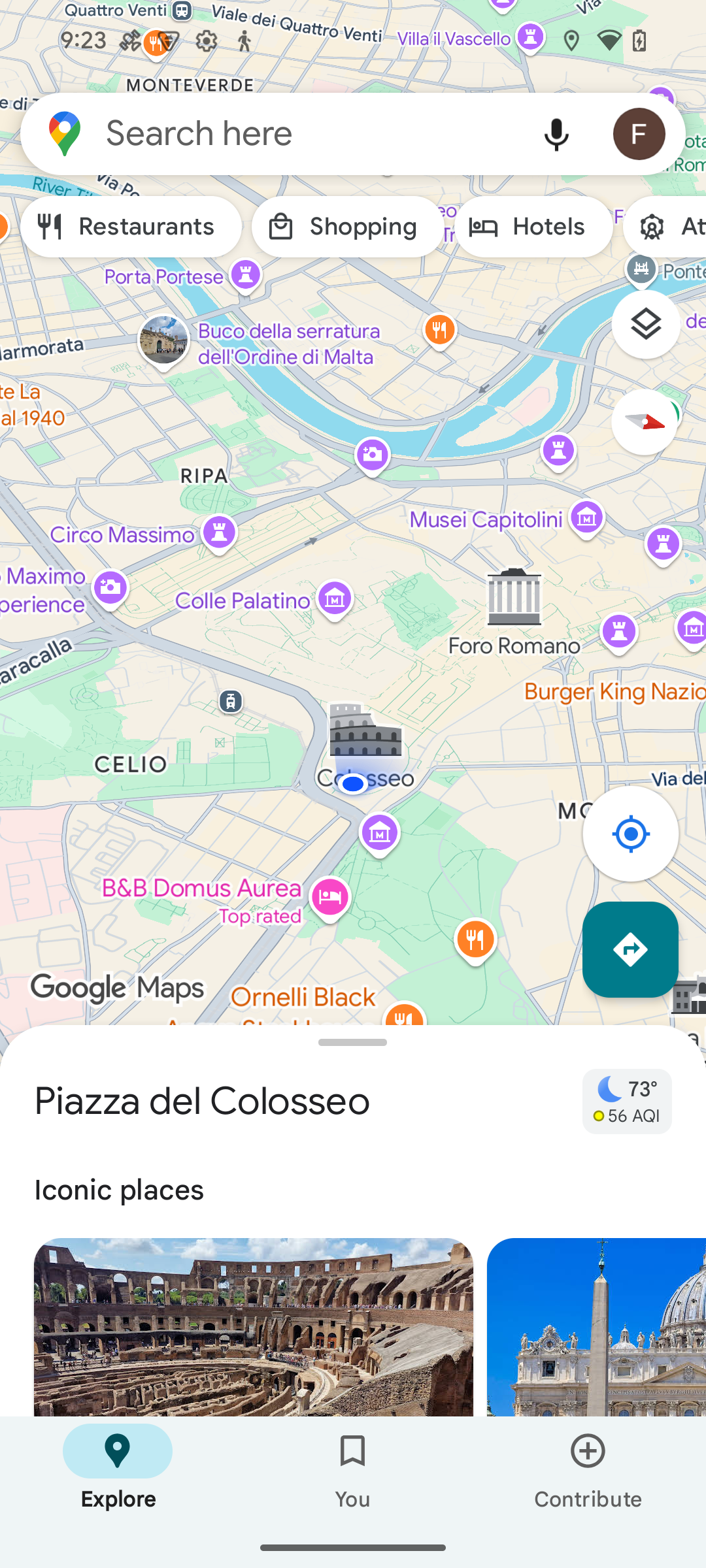}
        \captionsetup{font=footnotesize}
        \subcaption{GPS spoofed to Piazza del Colosseo, Rome. Location change triggers contextual adjustments across location-aware apps.}
        \label{fig:3c}
    \end{minipage}

    \vspace{1em}

    \begin{minipage}[t]{0.30\linewidth}
        \centering
        \includegraphics[height=0.40\textheight]{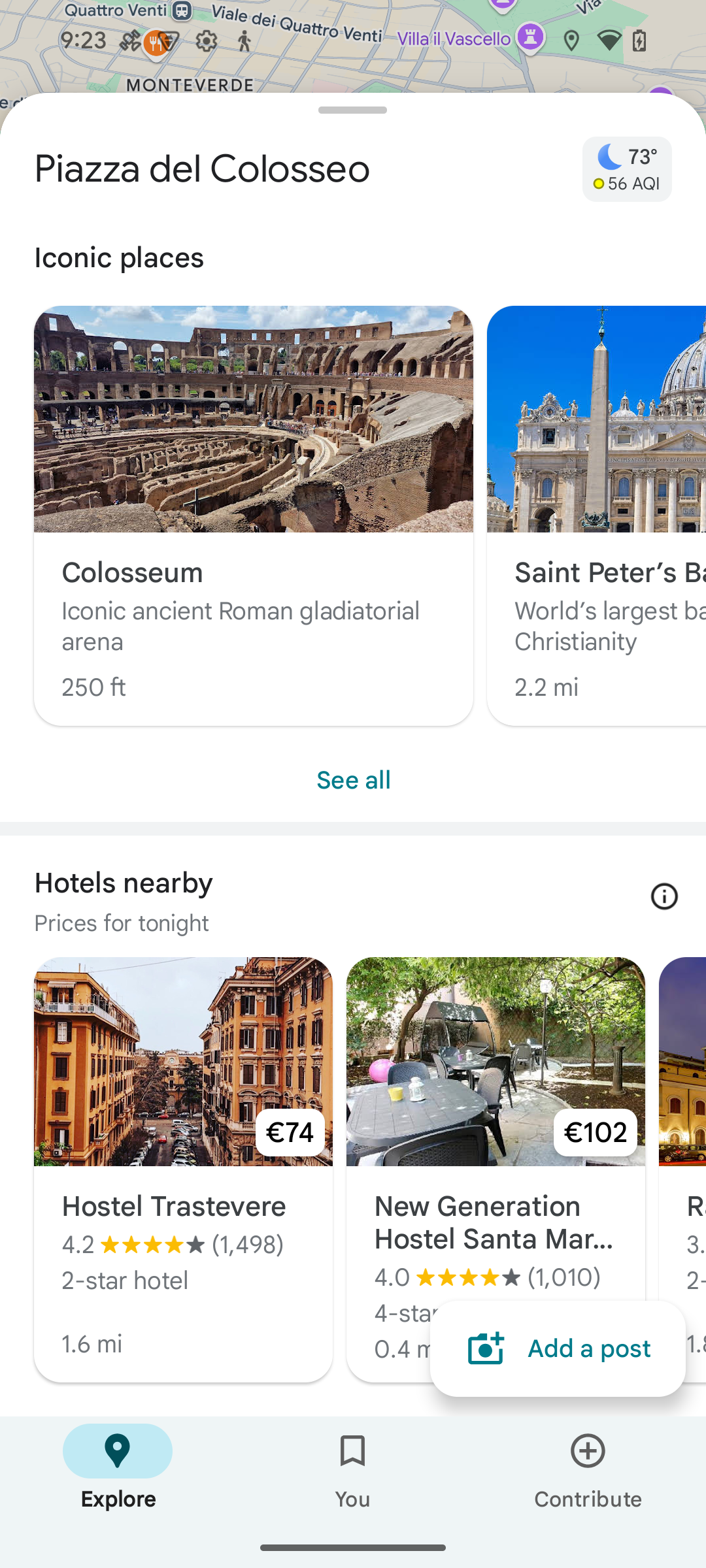}
        \captionsetup{font=footnotesize}
        \subcaption{Local discovery app displays "What's Nearby" recommendations tailored to Rome, verifying dynamic content adaptation to spoofed location.}
        \label{fig:3d}
    \end{minipage}
    \hfill
    \begin{minipage}[t]{0.30\linewidth}
        \centering
        \includegraphics[height=0.40\textheight]{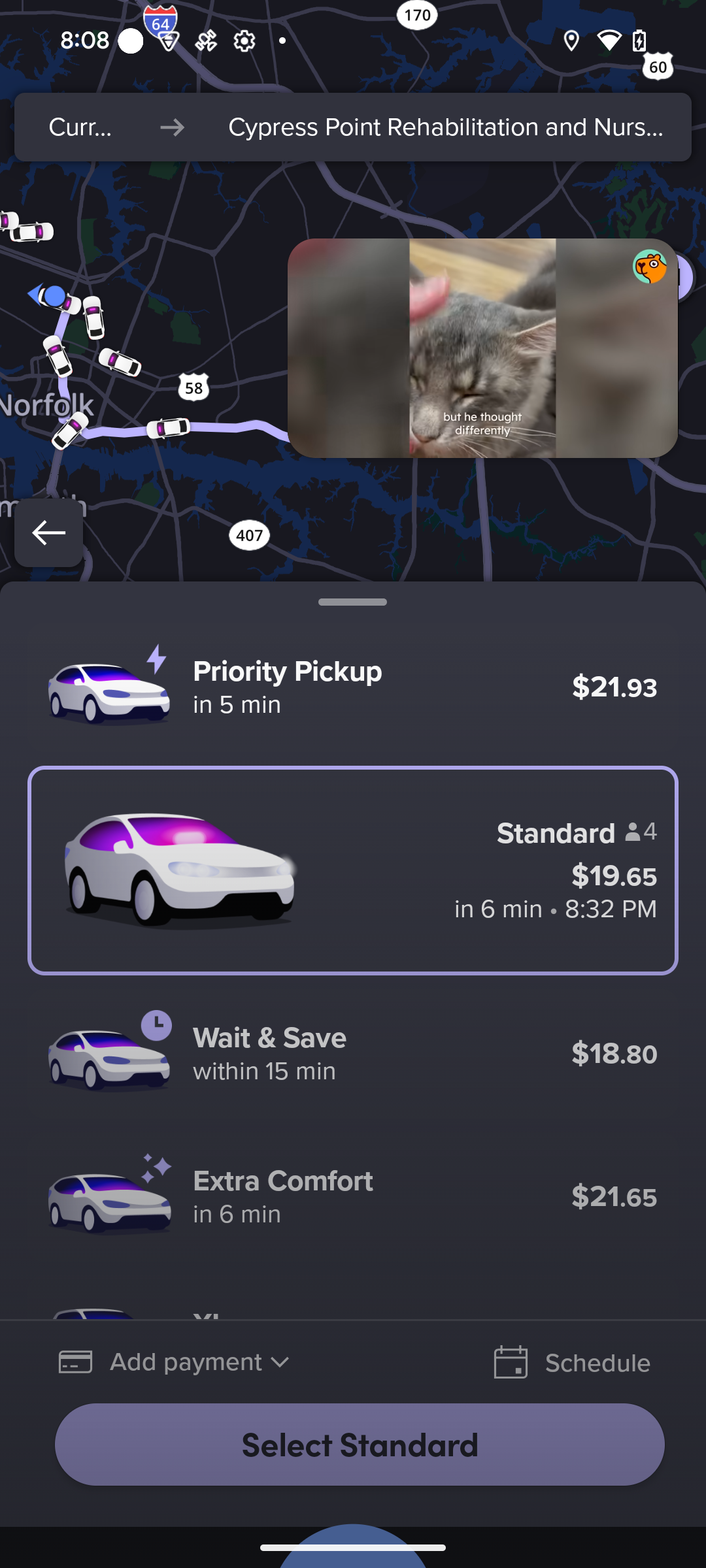}
        \captionsetup{font=footnotesize}
        \subcaption{Lyft app pricing shown in USD when system region and GPS indicate a U.S. location. Used as a baseline before spoofing.}
        \label{fig:3e}
    \end{minipage}
    \hfill
    \begin{minipage}[t]{0.30\linewidth}
        \centering
        \includegraphics[height=0.40\textheight]{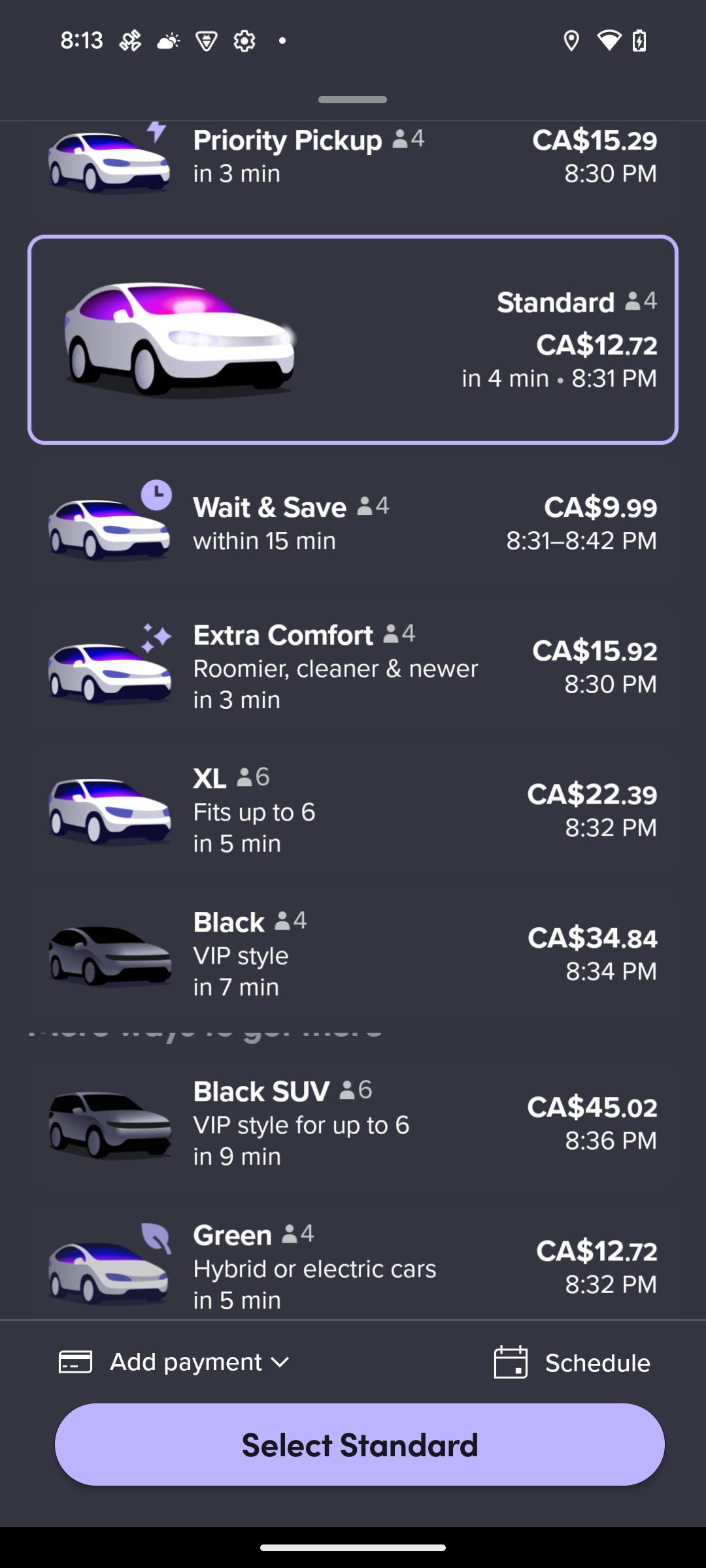}
        \captionsetup{font=footnotesize}
        \subcaption{Lyft app pricing automatically updates to CAD after GPS spoofed to Toronto, Canada—demonstrating region-sensitive pricing adaptation.}
        \label{fig:3f}
    \end{minipage}
     \caption{Preliminary results showing that mobile apps can respond to persona-driven sensor and location spoofing. These examples illustrate early signs of behavior adaptation—such as fitness badges, location-based content, and currency changes—and point toward the potential for detecting implicit data use patterns in future work.}
    \label{fig:3}
\end{figure*}

Fitness-focused apps displayed some of the most immediate and observable reactions. In the app \textit{Step Counter - Pedometer}, injecting high-frequency step counter values along with accelerometer drift led to rapid increases in step tallies. The app promptly responded with congratulatory pop-ups, motivational notifications, and goal-based achievement badges, even in the absence of any physical activity (see Figure ~\ref{fig:3a}, \ref{fig:3b}). This behavior illustrates that such apps rely directly on real-time sensor values and are quick to generate feedback loops based on those inputs.

Weather and utility apps showed similarly responsive behavior. When the device's GPS and system time were spoofed to simulate nighttime in a different city, the UI adapted accordingly, switching to night mode and updating forecasts for the spoofed region (Figure~\ref{fig:3c}). These changes confirm a tight coupling between ambient sensor signals and app interface logic.

In navigation and transportation apps, spoofing location data produced more nuanced effects. In the Lyft app, for instance, changing the GPS coordinates to a Canadian city resulted in fare estimates being displayed in CAD instead of USD (Figure~\ref{fig:3f}),  while a U.S. location showed USD (Figure~\ref{fig:3e}). More strikingly, setting the GPS to a country where Lyft does not operate (e.g., certain regions in Europe or Asia) triggered fallback messages indicating the service was unavailable in the spoofed location. These examples highlight that even core app functionality can dynamically shift based on geographic sensor input.

E-commerce platforms such as AliExpress were more conservative in their adaptation. While our system spoofed location and time to simulate browsing from Rome during night hours, the app interface did not automatically localize content based on GPS alone. Instead, region-based personalization appeared to depend on account-level settings or explicit region selection, suggesting the presence of internal gating logic before applying contextual changes. This contrast underscores the variability in how apps integrate sensor data into their personalization pipelines.

Importantly, not all applications responded uniformly. Some apps remained inert to spoofed inputs unless users interacted with specific features, while others showed delayed personalization effects, offering tailored suggestions or notifications hours after the spoofed conditions were applied. These varied behaviors suggest that some personalization mechanisms are event-triggered or tied to backend inference models that process composite behavioral patterns over time.

Our toolkit enables researchers and end-users to systematically surface these hidden behaviors without modifying app binaries or requiring advanced technical expertise. Although our tests represent early-stage evaluations, the results highlight the importance of multi-sensor awareness in privacy audits and suggest that further exploration, such as combining spoofed browsing history, long-term persona routines, or app engagement patterns, may uncover deeper layers of behavioral inference across mobile ecosystems.

\section{Discussion}

Our preliminary findings confirm that mobile apps dynamically adapt to behavioral signals spoofed via sensor data. These changes, ranging from fitness badges to localized content and UI shifts, demonstrate that sensor-driven personalization is both active and observable. Building on this, our next steps focus on strengthening the system’s technical capabilities and exploring how users interpret these behaviors.

In terms of system refinement, we are expanding spoofing beyond environmental sensors to include identity-linked traits such as browser history, calendar routines, and advertising IDs. This will help us examine how higher-level contextual signals influence content personalization, including ad delivery and UI behavior. Additionally, we plan to enhance instrumentation to log network activity, app logic, and sensor access over time, using tools like the Android Privacy Dashboard \cite{privacydashboard} to help map how and when apps access specific types of data.

On the user-facing side, we are developing a lightweight mobile interface that lets individuals select and activate personas (e.g., student, traveler, fitness enthusiast), each mapped to curated sensor profiles. This will support non-technical users in running their own spoofing sessions and observing app responses in real time.

To evaluate the system's impact, we are designing a small-scale user study in which we will present participants with persona-driven app experiences. Through think-aloud protocols and interviews, we aim to understand how users interpret content changes, whether they can identify sensor triggers, and whether these interactions foster greater privacy awareness or control.

After establishing the technical feasibility of persona-driven sensor spoofing, the next stage of this work turns toward the human dimension. The value of this system lies not only in its ability to elicit measurable behavioral changes from mobile apps, but also in revealing how people interpret these changes, how such interpretations shape their privacy attitudes, and whether they lead to different decision-making over time. Addressing these questions will involve conducting user studies in a controlled lab environment, where participants engage with persona-driven app experiences and subsequently take part in semi-structured interviews and open discussions. Both the system's behavioral logs and participants' qualitative responses will be analyzed in tandem, enabling us to link observed personalization changes with how users perceive and interpret them. This dual analysis will provide richer insight into the relationship between technical adaptation mechanisms and human privacy reasoning. By structuring the project in this way, we ensure that technical refinements directly enable richer, more realistic scenarios for human-centered inquiry, ultimately allowing the platform to serve as both a diagnostic tool for app behavior and a catalyst for meaningful discussion on mobile privacy and personalization.

Together, these efforts aim to reposition behavioral spoofing as a user-facing method for auditing mobile personalization. Rather than being a circumvention tactic, our approach frames spoofing as a form of transparency—empowering users to test, reflect on, and better understand how their mobile behaviors are interpreted and influence app experiences.

As the system matures, we also plan to situate its findings within the broader landscape of known personalization behaviors. Specifically, we will compare observed adaptations against patterns already documented in public sources, such as user reports, developer forums, and prior research, to further validate the novelty that our sandbox produces. This comparison will help us evaluate the added value of active persona-driven testing over passively collected or crowd-sourced observations, clarifying the kinds of insights our approach can uniquely provide.

\section{Conclusion}

This work introduces a novel sandbox-based toolkit that repositions sensor spoofing as a constructive, user-facing mechanism for transparency and auditing—rather than solely a tool for adversarial attack. By simulating behavioral contexts through persona-driven sensor inputs, our system enables users and researchers to visualize how mobile apps adapt in response to inferred identities and routines. These adaptations, though often invisible, shape content delivery, personalization pathways, and even user trust in ways that are seldom made transparent.

While still early in development, this approach offers a new direction for privacy-enhancing technologies: one grounded in active exploration and experiential awareness. Instead of shielding users from data flows through static protections, our system empowers them to interrogate and reflect on those flows interactively. The combination of real-time sensor spoofing, automated persona generation, and visual UI analysis not only supports empirical studies of app behavior but also suggests a future in which mobile privacy tools can foster critical digital literacy.

By making app personalization visible and testable through real-time sensor manipulation, our system helps shift privacy tools away from passive restriction and toward active engagement. Instead of simply limiting data access, users are given the opportunity to explore how different behaviors influence app responses. This visibility encourages deeper understanding of personalization mechanisms and can support future efforts in privacy education, design evaluation, and user research. When users can see how routine actions—like commuting patterns or late-night browsing—shape their digital environment, they are better equipped to reflect on the trade-offs they make in everyday app use.

We call on researchers, developers, and platform designers to consider sensor spoofing as a valuable auditing strategy—capable of surfacing opaque inference mechanisms and informing the design of more transparent and accountable personalization systems. Future iterations of this work may integrate feedback loops, participatory design frameworks, or even crowd-sourced persona libraries to better align with diverse user needs and ethical considerations.

As mobile ecosystems continue to deepen their reliance on behavioral sensing, systems like ours can help ensure that users are not merely passive recipients of personalization, but active participants in shaping the terms of their digital experiences. Equipping individuals with tools to explore and understand how personalization systems operate can foster greater transparency, user agency, and critical awareness in today’s data-driven environments.

\begin{acks}
    The authors would like to thank the anonymous reviewers for their insightful feedback. This research is in part supported by the National Science Foundation CNS-2341187, CNS-2426397, CNS-2442221, CNS-2426395, CCF-2211428, CMMI-2326378, CNS-2426396, a Google PSS Faculty Award, and a Meta Research Award. Any opinions, findings, and conclusions or recommendations expressed in this material are those of the authors and do not necessarily reflect the views of the sponsors.
\end{acks}

\bibliographystyle{ACM-Reference-Format}
\bibliography{bibliography}

\end{document}